\documentclass[11pt]{article}
\usepackage{moriond,epsfig,wrapfig,amsmath}

\def\Journal#1#2#3#4{{#1} {\bf #2}, #3 (#4)}

\def\MNRAS{\em MNRAS}
\def\APJ{\em ApJ}

\def\mco{\multicolumn}

\begin{document}
\vspace*{4cm}
\title{MAXIMUM--ENTROPY RECONSTRUCTION OF THE DISTRIBUTION OF MASS IN
CLUSTER MS1054-03 FROM WEAK LENSING DATA}

\author{ P.J. Marshall }

\address{Astrophysics Group, Cavendish Laboratory, Madingley Road,
Cambridge CB3 0HE, England}

\maketitle\abstracts{Weak gravitational lensing studies of clusters of 
galaxies provide
complimentary information to that from X-ray data; the measured signal
depends only on the cluster's projected mass, independent of dynamical
assumptions.   
Here we apply a maximum--entropy method algorithm for reconstructing 
the two-dimensional density
distribution\cite{bhls} to two sets of shear data for
the high redshift cluster MS1054-03, courtesy of
Clowe et~al (2000)\cite{clo} and Hoekstra et~al (2000).\cite{hfk}}

\section{Weak Lensing Introduction}

The mass in a cluster of galaxies produces a net distortion of the
shapes of images of galaxies lying behind the cluster. The galaxies
appear stretched tangentially around the mass concentrations. 
This distortion is often
described by the \emph{reduced shear} field~$g$:
\begin{equation}
 g = \frac{\gamma}{1-\kappa}
 \notag
\end{equation}
\noindent where the \emph{convergence}~$\kappa$ is the projected mass density of 
the cluster relative to
a critical density
$\Sigma_{crit}$ (which is a function of
$z_{cluster},z_{galaxies},\Omega_m,\Omega_{\Lambda}$ and $H_0$).
The \emph{shear}~$\gamma$ is related to the convergence by a
convolution over the whole image plane:

\begin{wrapfigure}{r}{6cm}
\begin{center}
  \epsfig{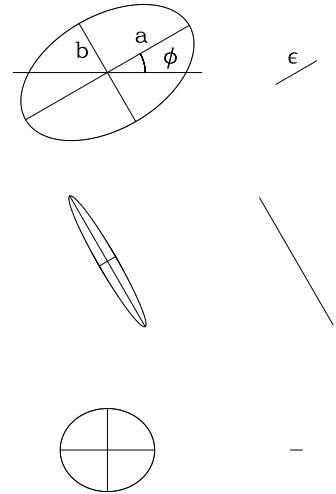}
\end{center}
\caption{3 model galaxy images, and their representations in terms of the 
ellipticity parameters described in the text.}
\end{wrapfigure}
\begin{equation}
 \gamma (\theta) = \frac{1}{\pi} \int D(\theta - \theta')
          \kappa(\theta') d^2 \theta'
 \notag
\end{equation}
\noindent If we parameterise the
(elliptical) galaxy image shapes by

\begin{equation}
 \epsilon = \epsilon_1 + i\epsilon_2 = 
  \biggl( \frac{1-\frac{b}{a}}{1+\frac{b}{a}} \biggr) e^{2i\phi}
 \notag
\end{equation}

\noindent then averaging over many galaxies (to remove the intrinsic 
shapes) can be shown to give an unbiased estimate of the reduced shear:\cite{ks}
\begin{equation}
 \left< \epsilon \right> = g
 \notag
\end{equation}
\noindent We want to infer $\kappa$ from a measured $g$ field...

\clearpage

\section{Maximum Entropy Weak Lens Reconstruction}

\noindent We aim to find the convergence map that maximises the
posterior probability:\cite{bhls}
\begin{equation}
  Pr(\kappa | data) = \frac{Pr(data | \kappa) Pr(\kappa)}{Pr(data)}
\notag
\end{equation}
\noindent The data are the measured values of~$g_1$ and $g_2$ in the
pixels of the background galaxy averaging grid; $\kappa$ is
reconstructed on a larger grid, since mass lying outside the observing
region will produce a shear signal within it.

\vspace{\baselineskip}
\noindent Assuming Gaussian errors on the~$g$ values, the likelihood is
\begin{equation}
  Pr(data | \kappa) \propto e^{-\frac{\chi^2}{2}}
\notag
\end{equation}
\begin{equation}
  \text{where   } \chi^2 = \sum_i \frac{(g_i^{observed} -
  g_i^{predicted})^2}{\sigma_i^2}
\notag
\end{equation}
\noindent The prior probability has to be specified -- an 
\emph{entropic} prior is 
appropriate:
\begin{equation}
  Pr(\kappa) \propto e^{\alpha S}
\notag
\end{equation}
The entropy~$S(\kappa)$ becomes more negative as~$\kappa$ departs 
from the default 
model --
taken to be~$\sim$zero mass density; it acts to suppress over-fitting
to the noise in the data, and to control the reconstruction of mass
outside the observed field where the constraints from the data are weak. 
Minimising $F=\chi^2 / 2 - \alpha S$ gives
the best-fit convergence distribution~$\hat{\kappa}$.

\vspace{\baselineskip}
\noindent The uncertainties in~$\hat{\kappa}$ are estimated by
approximating~$Pr(\kappa | data)$ by a multivariate Gaussian:
\begin{equation}
  Pr(\kappa | data) \approx \exp \left[ -\frac{1}{2} (\kappa -
  \hat{\kappa})^T \nabla_{\kappa} \nabla_{\kappa} F (\kappa - \hat{\kappa})
  \right]
\notag
\end{equation}
\noindent So we have the covariance matrix for the convergence 
pixel values, the (square root of the) diagonal elements of which 
provide an estimate of the uncertainty
in each pixel value.

\vspace{\baselineskip}
\noindent To produce the plots in this poster the reconstructed
convergence distributions have been smoothed; the ``signal to noise
contours'' were obtained by dividing the reconstructed convergence in
each pixel by its uncertainty, and then smoothing on the same scale.

\section{The MS1054-03 Data}

MS1054-03 is a high redshift ($z=0.83$) galaxy cluster; X-ray
and dynamical measurements suggest a high mass 
($T_X \approx 12.3 keV$,\cite{don}, $\sigma \approx 1150$
km~s$^{-1}$\cite{vdokk}).
2 sets of weak lensing data have been analysed:

\begin{itemize}
 \item{Clowe et al 2000} -- Single Keck pointing 
  $\rightarrow$~$\epsilon$ measured for 2723 background galaxies in a
  50 square arcminutes field
 \item{Hoekstra et al 2000} -- Irregularly shaped mosaic of 
  6 HST WFPC2 images 
  $\rightarrow$~$\epsilon \pm \delta \epsilon$ measured for 
  2446 background galaxies (in a region of area $\sim 30$ arcmin$^2$).
  A higher density of galaxy images combined
  with a much smaller point spread function means that the data is
  of higher
  quality -- a higher resolution reconstruction can be performed.
\end{itemize}  

\begin{figure}[!ht]
\begin{minipage}[t]{0.48\linewidth}
\centering\epsfig{file=figure2a.eps,height=7cm,angle=270,clip=}
\end{minipage} \hfill
\begin{minipage}[b]{0.48\linewidth}
\centering\epsfig{file=figure2b.eps,height=7cm,angle=270,clip=}
\end{minipage}
\end{figure}

\begin{figure}[!ht]
\begin{minipage}[t]{0.48\linewidth}
\centering\epsfig{file=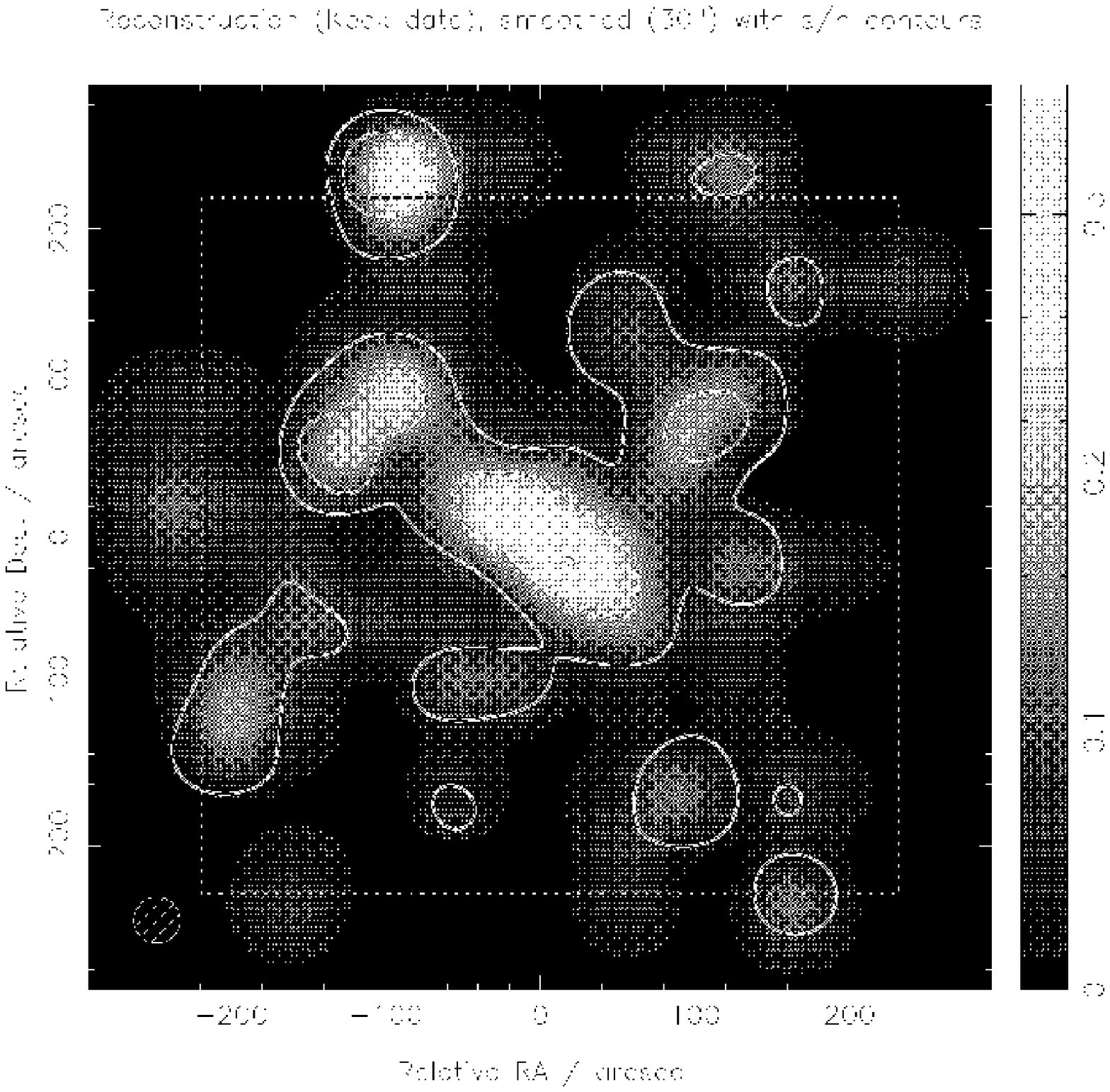,height=7.5cm,angle=270,clip=}
\end{minipage} \hfill
\begin{minipage}[b]{0.48\linewidth}
\centering\epsfig{file=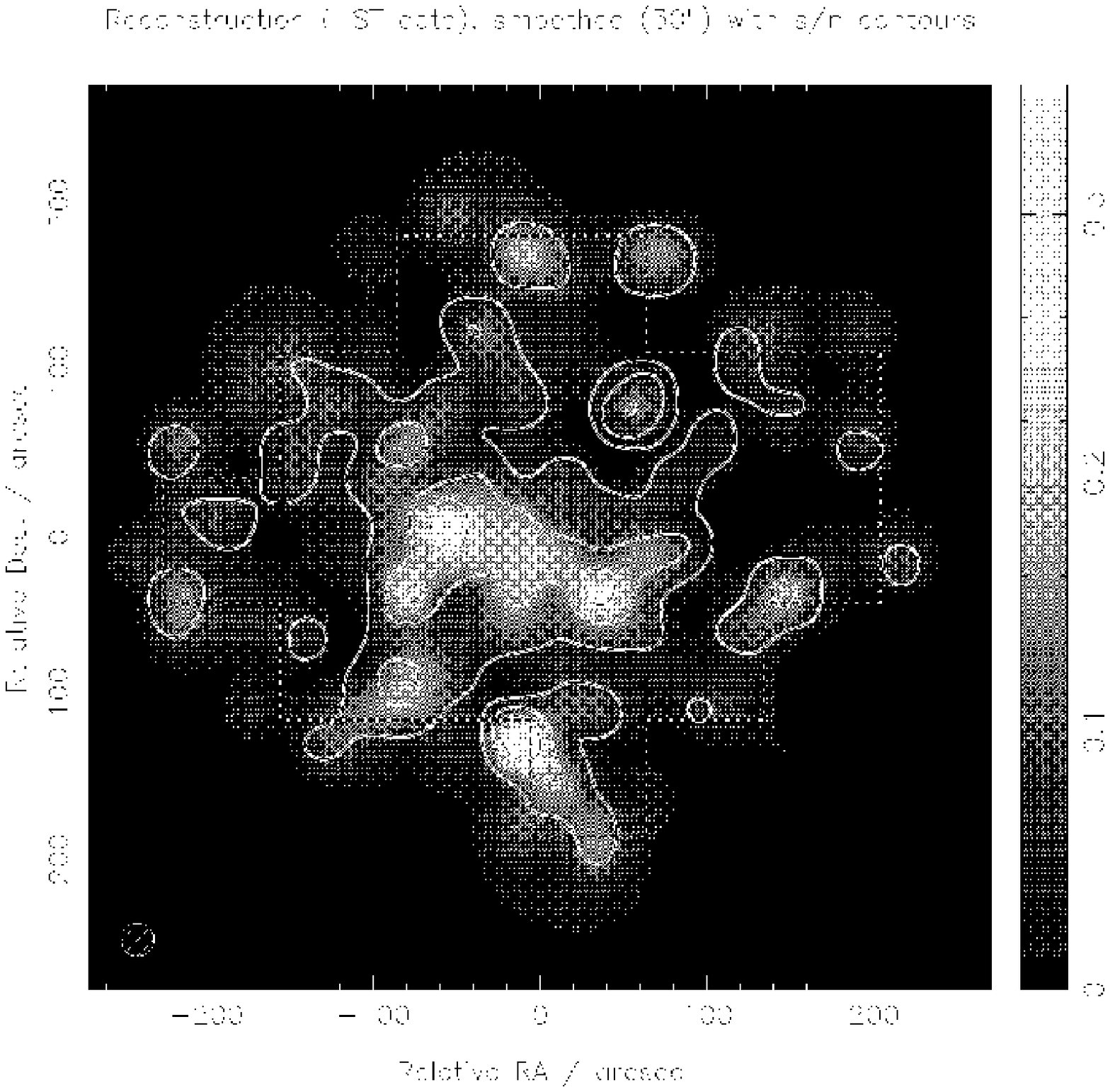,height=7.5cm,angle=270,clip=}
\end{minipage}
\end{figure}

\section{The LensEnt Reconstructions}

\begin{itemize}
 \item{Keck data} -- The cluster is visible as 
 the central North-East/South-West elongated
 blob; the highest contour is 3$\sigma$. Other
 features are present, possibly due to noise. 
 \item{HST data} -- The shape of the cluster is
 shown to apparently higher resolution, but again there are 
 many noise peaks.
 As for the
 cluster, how many sub-clumps does it contain? 1, 2 or 3? 
\end{itemize}  

\enlargethispage{\baselineskip}

\subsection{How Much Structure Do We Believe?}

The plotted signal-to-noise contours show the significance 
levels of structure if we accept
both the data and the reconstruction at
face value -- if present, outliers in the data could 
introduce spurious signals.

\vspace{\baselineskip}
\noindent As a test,\cite{hfk} the shears were rotated
by~$\frac{\pi}{4}$ to reconstruct the 
``imaginary'' convergence -- the cluster (having real mass!) 
disappears
but spurious features in the North (Keck data) and
South-East (HST data) remain, 
presumably due to noise spikes in the data.

\begin{figure}[!ht]
\begin{minipage}[b]{0.48\linewidth}
\centering\epsfig{file=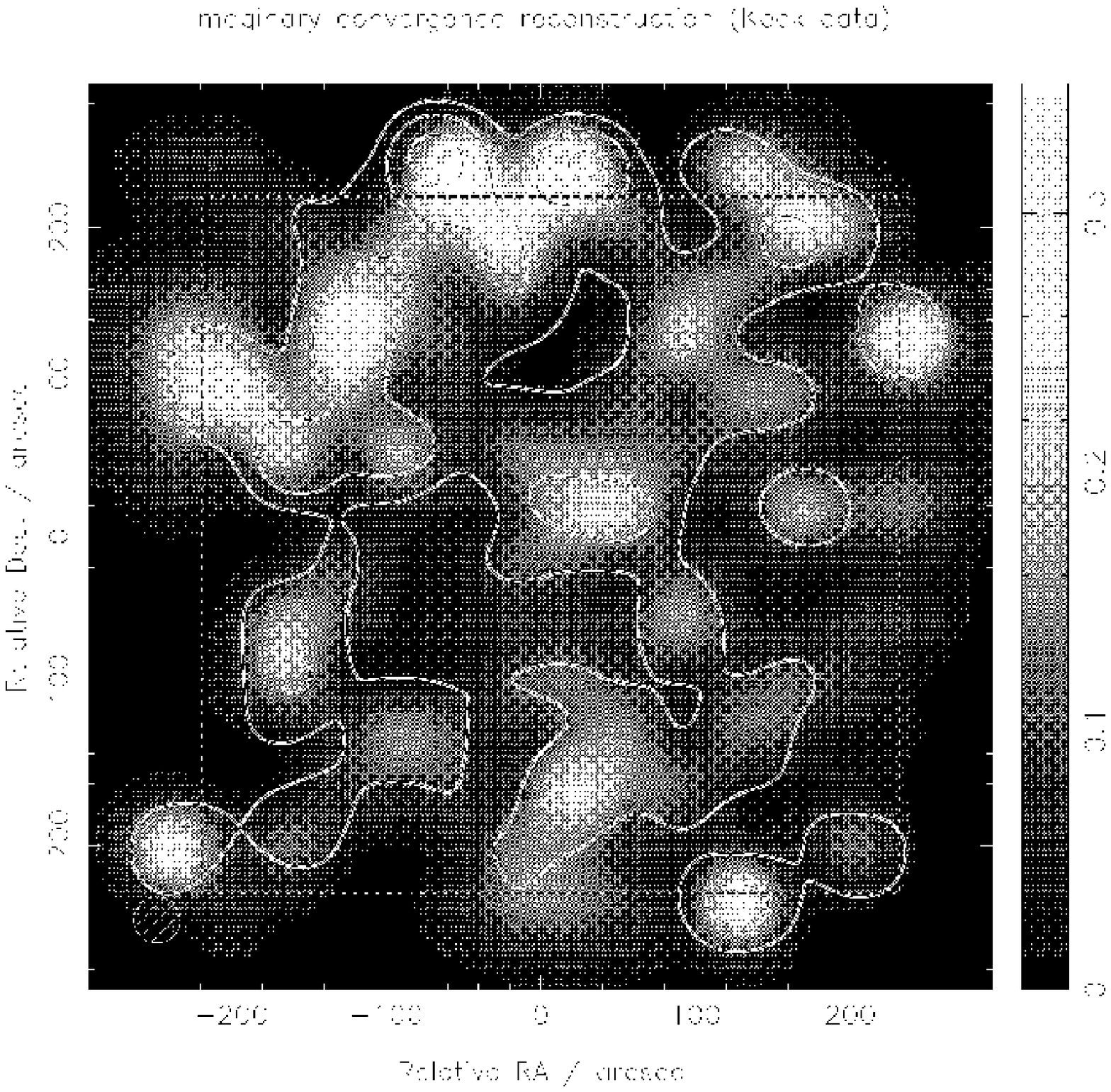,height=7.5cm,angle=270,clip=}
\end{minipage} \hfill
\begin{minipage}[b]{0.48\linewidth}
\centering\epsfig{file=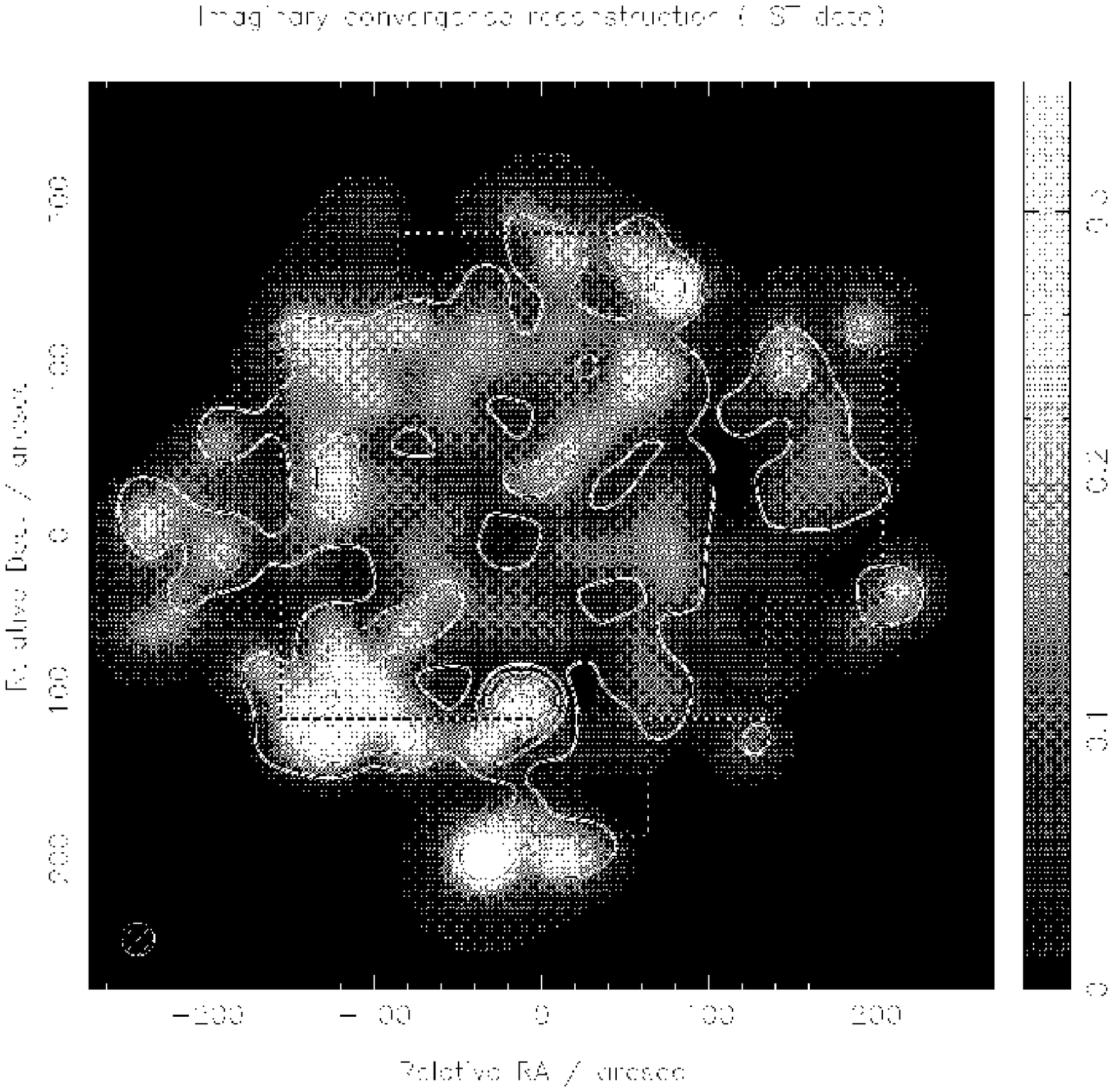,height=7.5cm,angle=270,clip=}
\end{minipage}
\end{figure}

\noindent We can illustrate our new knowledge of the cluster's shape by
plotting samples drawn from the posterior probability distribution --
approximately $\frac{2}{3}$ of the sample images will lie 
within~$\pm 1
\sigma$ of the best-fit reconstruction. A wide range of structure is
consistent with the shear data!

\begin{figure}[!b]
\centering\epsfig{file=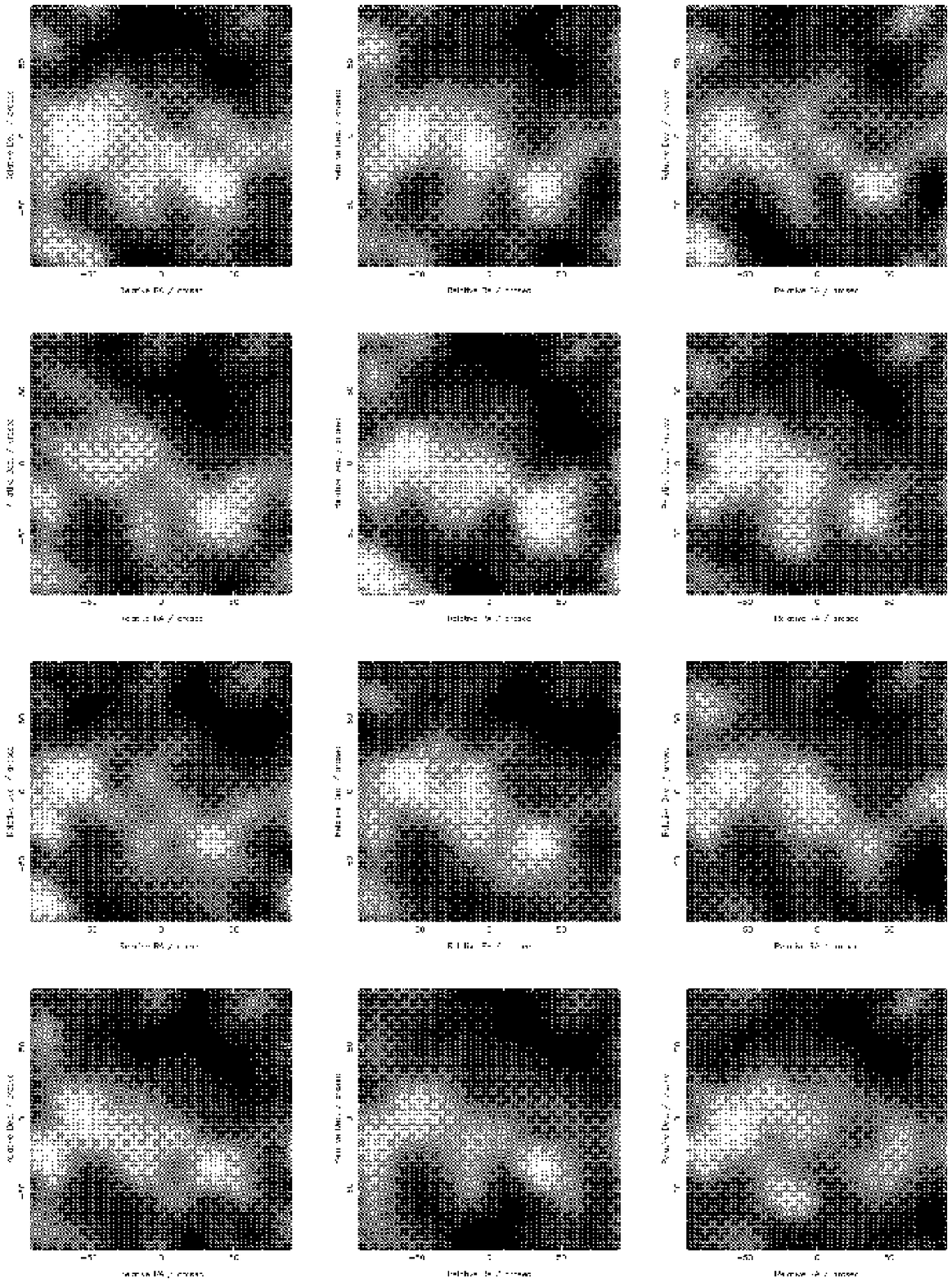,height=13cm,angle=0,clip=}
\end{figure}

\clearpage

\subsection{Mass Estimation}

Given the reconstructed convergence
distribution~$\hat{\kappa_i}$ 
(with
associated covariance matrix $C_{ij}$), and a value
for
$\Sigma_{crit}$ (calculated by Hoekstra et al\cite{hfk} and so only 
really
appropriate for that dataset),
we can estimate the projected mass within a circular aperture:

\begin{xalignat}{2}
 M &= \sum_i a_i \kappa_i & \pm \; \sigma_M &= \sum_{ij} a_i a_j C_{ij}
 \notag
\end{xalignat}  

\noindent where $a_i$ is the proportion of the area of the $i^{th}$ 
pixel lying within the aperture. 

\begin{figure}[!h]
\begin{minipage}[!t]{0.48\linewidth}
\centering\epsfig{file=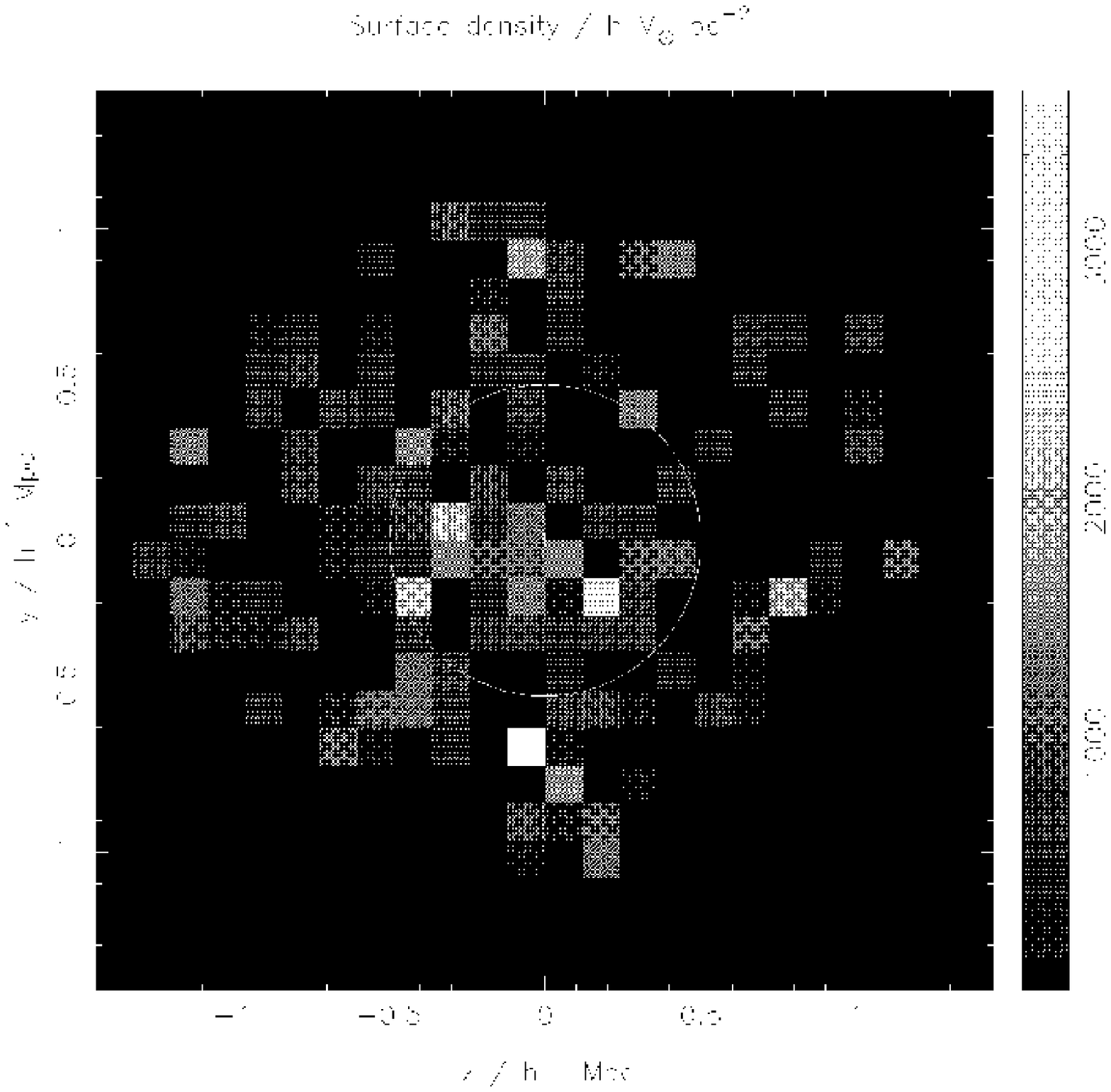,width=\linewidth,angle=270,clip=}
\end{minipage} \hfill
\begin{minipage}[!t]{0.48\linewidth}
\begin{tabular}{|cccc|}
\hline
           &                    &                   &                               \\
           &                    &     HST data      &     Keck data                 \\
$\Omega_m$ & $\Omega_{\Lambda}$ & \mco{2}{c|}{$M(<\,1\,h_{50}^{-1}\,Mpc)\;/\;10^{15}\,h_{50}^{-1}\,M_{\odot}$} \\ 
\hline
           &                    &                   &                               \\ 
0.3        & 0.0                & $1.11\pm0.09$     & $1.33\pm0.15$                 \\
0.3        & 0.7                & $0.89\pm0.07$     & $1.14\pm0.12$                 \\
           &                    &                   &                               \\ 
\hline
\end{tabular}
\end{minipage}
\end{figure}

\section{Is This The Best We Can Do?}

So far we have assumed that the convergences in each pixel are
independent; this is not the case for physical cluster structure
on larger scales than one pixel.
An intrinsic correlation 
function~(ICF),\cite{rob}
that maps some ``hidden'' convergence distribution (which has an
entropic prior) onto the ``visible'' distribution used to fit the data,
eases the introduction of large-scale structure into the
reconstruction. One step further is to use a multiple-scale ICF
(still under development). The reconstruction from the HST data is
smoother, with the cluster appearing as a possible two-component 
extended feature.
This map has been centred on the CD galaxy, with North upwards and East
to the left. Contours are signal-to-noise as before.

\vspace{\baselineskip}
\noindent The inclusion of an ICF does, to some extent, add unwanted 
smoothness
-- a small scale convergence distribution (fed back from the
previous results) is reconstructed with some
increase in
smoothness. The
converse is also true: the smooth output from the multi-resolution ICF 
reconstruction is
made ``grainier'' by the single scale reconstruction code in the
presence of high noise. So which reconstruction should we prefer? The one that
maximises the probability of getting the data, given a hypothesis of
the degree of smoothness of
the convergence distribution -- the EVIDENCE, $Pr(data)$.
This does indeed turn out to be higher for the multi-scale
ICF reconstruction. 

\begin{figure}[!ht]
\begin{minipage}[!t]{0.48\linewidth}
\centering\epsfig{file=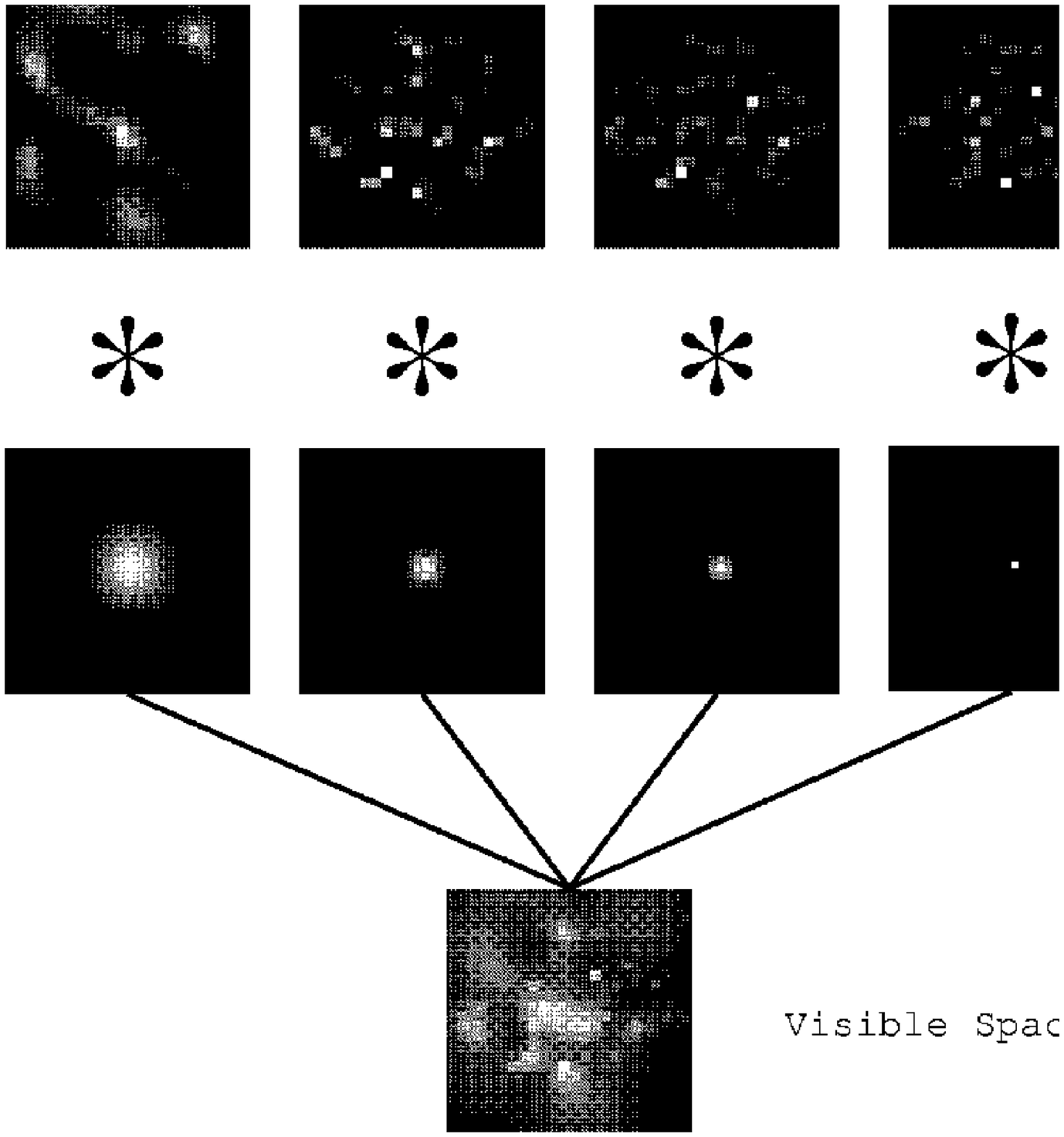,width=\linewidth,angle=0,clip=}
\end{minipage} \hfill
\begin{minipage}[!t]{0.48\linewidth}
\centering\epsfig{file=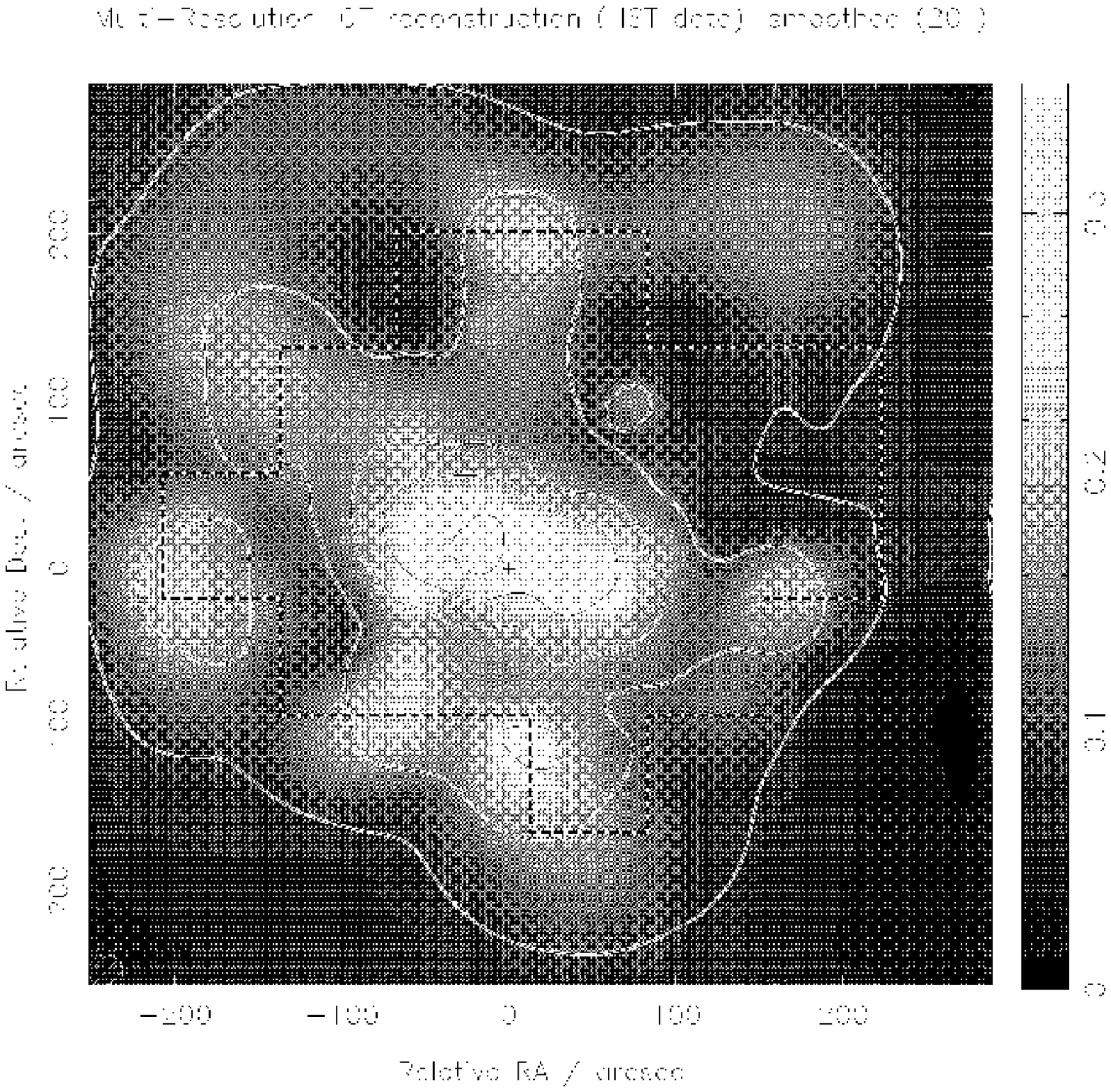,width=\linewidth,angle=270,clip=}
\end{minipage}
\end{figure}

\section{Conclusions}

\begin{itemize}
 \item Reconstruction of the mass distribution in MS1054-03 from both 
       Keck and 
       HST data show the non--spherical cluster shape found by previous
	 authors.  
 \item Evaluation of the reconstruction uncertainties show that the
       cluster substructure is present at relatively low significance;
	 sample convergence maps show a wide range of possible cluster
	 configurations consistent with the weak lensing data.  
 \item Given the (small) statistical and (larger) systematic errors 
       involved, 
       a mass estimate of 
	 $1 \times 10^{15} h_{50}^{-1} M_{\odot}$ would be a reasonable
	 consensus value for MS1054-03 -- as found by Hoekstra et al. 	    
 \item The inclusion of an ICF in the algorithm appears promising,
       particularly with regard to determining the level of substructure
	 really present in clusters of galaxies; much more investigation
	 is required before drawing any definite conclusions!   
\end{itemize}  
  
\noindent The LensEnt reconstruction code used for the first part of this
analysis is available from 

\vspace{\baselineskip}
\texttt{http://www.mrao.cam.ac.uk/projects/lensent/}
\vspace{\baselineskip}

\noindent with the ICF extensions to follow in the very near future.

\end{document}